\documentclass[10 pt]{iopart} 

\begin{document}        

\title{Structure and Parameterization of Stochastic Maps of Density Matrices}

\author{E C G  Sudarshan\dag\ and Anil Shaji\ddag}
 
\address{\dag\ Center for Particle Physics, Department of Physics, University of Texas, Austin, Texas 78712}

\address{\ddag\ Center for Statistical Mechanics, Department of Physics, University of Texas, Austin, Texas 78712}

\ead{shaji@physics.utexas.edu}

\begin{abstract}
The most general evolution of the density matrix of a quantum system with a finite-dimensional state space is by stochastic maps which take a density matrix linearly into the set of density matrices. These dynamical stochastic maps form a linear convex set that may be viewed as supermatrices. The property of hermiticity of density matrices renders an associated supermatrix hermitian and hence diagonalizable. The positivity of the density matrix does not make the associated supermatrix positive though. If the map itself is positive, it is called completely positive and they have a simple parameterization. This is extended to all positive (not completely positive) maps. A general dynamical map that does not preserve the norm of the density matrices it acts on can be thought of as the contraction of a norm-preserving map of an extended system. The reconstruction of such extended dynamics is also given. 

\end{abstract}

\submitto{\JPA}

\maketitle

\section{Introduction}
A quantum system with a finitely many dimensional state space may be represented by a $N \times N$ quantum density matrix $\rho$. The density matrix must be of trace class and should satisfy the properties of hermiticity and  positivity:
\begin{equation}
{\mbox{tr}}(\rho) = 1 \quad ; \quad \rho^{\dagger} = \rho \quad ; \quad x_r^* \rho_{rs} x_s \geq 0 .\label{constraints}
\end{equation}
For a closed system the dynamical evolution of the system is by the action of a unitary time-dependent operator.
\begin{equation}
\rho(t_2) = U(t_1, t_2) \rho(t_1) U^{\dagger}(t_1, t_2) \label{unitary}
\end{equation}
where
\[ U(t_1, t_2) = {\cal{T}} \left\{ \exp \left( - i \int_{t_1}^{t_2} H(t') dt' \right) \right\}. \]
The evolution is linear. But if we have an open system and if we are considering the back reaction of the environment on the system then the dynamics cannot be by unitary evolution but by a more general linear evolution\cite{ecg1}:
\begin{equation}
\rho(t_2) = A(t_1, t_2) \rho(t_1). 
\end{equation}
The linearity of $A(t_1, t_2)$ follows from the linearity of quantum mechanics and for this reason we do not consider more complicated forms of maps on density matrices.  The superoperator $A$ can be written as a supermatrix and the transformation can be written as 
\[ \rho_{rs} \longrightarrow A_{rs;r's'} \rho_{r's'} = (A\rho)_{rs} .\]
In the equation given above, the elements of the density matrix has been suitably regrouped into a column vector so that $A$ can be in the form of a $N^2 \times N^2$ matrix. The constraints (\ref{constraints}) on the density matrix impose restrictions on $A$. It is instructive to first recast $A$ into another dynamical matrix \cite{choi, choi1, davies, stromer} $B$ such that
\[ A_{rs; r's'}(t) = B_{rr'; s's}(t) .\]
In the form $B$ the supermatrix has to satisfy the following relations:
\begin{eqnarray}
 B^*_{s's;rr'}(t) \; =& B_{rr';s's}(t)  &{\mbox{(Hermiticity)}} \\
B_{nr'; s'n} \; =& \delta_{r's'}  \quad \quad &{\mbox{(Normalization)}} \\
x^*_r y_{r'} B_{rr';s's} x_s y^*_{s'} \; \geq& 0 \qquad \quad  &{\mbox{(Positivity)}}. \label{positivity}
\end{eqnarray}
In terms of the pairs of indices $rr'$ and $s's$, $B$ is a hermitian matrix which gives non-negative expectation values for factorizable supervectors.
\[ u^{\dagger} B u \geq 0 \quad {\mbox{if}} \quad u_{rs} = x_r y_s^* .\]
It is not necessary that $B \geq 0$ for maintaining the positivity of the density matrices under dynamical evolution even though it is a sufficient condition. If $B \geq 0$ we will call the map ``completely positive''\cite{choi}. The terminology is slightly confusing unless one keeps in mind that positivity of the map is a statement about its action on density matrices while complete positivity can be regarded as a statement about the map itself, in addition to saying something about its action. For instance the action of taking the transpose of a density matrix: $\rho \rightarrow \rho^T$ is a positive map but not a completely positive map. Complete positivity ensures that the evolution of the system of interest, $S$ due to the action of the map is extensible in a trivial way to a {\em physical} evolution in a larger system $S' \, \otimes \, S$. Such an extension is especially useful when $S$ is entangled to $S'$. 

Completely positive maps are interesting in the context of quantum information theory and quantum computing in that it provides a description of the types of dynamical evolution of a composite quantum system that can change the degree of entanglement between parts of the system. In other words we can deal with the open evolution of quantum systems rather than unitary isolated evolution thereby getting a handle on the difficult problem of the influence of the environment on a potential quantum computer. Extensive literature exists on the subject; we refer the reader to \cite{kraus, breuer} and the references therein. Completely positive maps have also been used in certain cosmological models and in attempts to construct a quantum theory of gravity in the context of Quantum Causal Histories \cite{fotini}.

In this paper we look at maps which are positive but not completely positive. Not completely positive maps do not represent dynamical evolution in the sense that completely positive maps do. This point is discussed in sections \ref{sec3} and \ref{sec4}.

The density matrices on which the dynamical maps act form a convex compact set. Since positive maps transform such a set into itself, they themselves form a convex set (the set of completely positive maps also form a convex set lying inside the set of positive maps). The convexity property means that any linear combination of of positive maps with nonnegative coefficients which sum to unity is also another valid  map. For instance the positive map $B$ defined by
\[ B = \sum_n k(n)B_n \quad ; \quad k(n) \geq 0 \quad ; \quad \sum_n k(n) =1 \] is also a positive map if $B_n$ are positive maps. It can also be shown that \cite{choi}\cite{ecg2} $1 \leq n \leq N^2$ where $N$ is the dimensionality of the density matrices that the map acts on. Out of the convex set of positive maps we can pick out those maps which cannot be written as a sum of other maps. Such maps are called extremal. In the discussion that follows we talk only about extremal maps and '$B$' is assumed to denote such maps. Any generic map can be constructed out of the extremal maps as a linear sum through the specification of at most $N^2$ extra parameters. We look for parameterizations of extremal maps. This, in conjunction with the $N^2$ extra parameters which add up to unity, will then suffice to parameterize a generic dynamical map acting on density matrices. Note that such a characterization of the extremal maps is usually carried out with the assumption that the additional restriction in equation (5) that the maps be trace preserving also holds. This restricts the maps being considered to a bounded convex set rather than to a convex cone.
 
Since $B$ is hermitian it follows that it has an eigenvector decomposition
\begin{equation}
B_{rr';s's} = \sum_{\alpha} \mu_{\alpha} \zeta_{rr'}^{(\alpha)} \zeta_{s's}^{(\alpha)*} = \zeta M \zeta^{\dagger},
\end{equation}
where $M$ is a diagonal matrix with eigenvalues $\mu_{\alpha}$ and $\zeta_{rr'}^{(\alpha)}$ are the normalized eigenvectors. For a completely positive map all the $\mu_{\alpha}$ are nonnegative but this is not true for all positive maps. If all the $\mu_{\alpha}$ are nonnegative, we can absorb them into the eigenvectors by defining
\[ C_{rr'}^{(\alpha)} =  \sqrt{\mu_{\alpha}}\;  \zeta_{rr'}^{(\alpha)} \]
So for a completely positive map\cite{choi}
\begin{equation}
 \rho \longrightarrow \sum_{\alpha} C^{(\alpha)} \rho C^{(\alpha) \dagger} \label{completely}
\end{equation}
with the trace condition
\[ \sum_{\alpha}  C^{(\alpha) \dagger} C^{(\alpha)} = 1 .\]
Parameterization of extremal completely positive maps can be found in reference \cite{ecg3}. Note that if the completely positive map is not extremal, its action can be written as
\begin{equation}
\rho \longrightarrow \sum_{n, \alpha} k(n)  C^{(\alpha)}_n \rho C^{(\alpha) \dagger}_n. \label{notextremal}
\end{equation}

\section{Not Completely Positive Maps}{\label{sec2}}

The action of an extremal map which is not completely positive on a density matrix can be expressed as 
\begin{equation}
\rho \longrightarrow \sum_{\alpha=1}^m C^{(\alpha)} \rho C^{(\alpha)\dagger} - \sum_{\beta=1}^n D^{(\beta)} \rho D^{(\beta)\dagger} 
\end{equation}
with
\[ D_{rr'}^{(\beta)} = (|\nu|)^{1/2}\eta_{rr'}^{(\beta)} \]
where $\nu$ are the negative eigenvalues of the map and $\eta_{rr'}^{(\beta)}$ are the corresponding eigenvectors.

It will turn out that the number of positive eigenvalues $m$ have to be greater than or equal to the number of negative eigenvalues $n$. The trace condition now becomes 
\begin{equation}
\sum_{\alpha=1}^m C^{(\alpha)\dagger} C^{(\alpha)} - \sum_{\beta=1}^n D^{(\beta)\dagger}  D^{(\beta)} = 1_{N \times N}.  \label{trace}
\end{equation}
The positivity condition yields the general result
\begin{equation}
\sum_{\alpha=1}^m C^{(\alpha)\dagger} u_{\alpha}u_{\alpha}^{\dagger} C^{(\alpha)} - \sum_{\beta=1}^n D^{(\beta)\dagger} u_{\beta}u_{\beta}^{\dagger} D^{(\beta)} \geq 0 
\end{equation}
Taking the trace on both sides of equation (\ref{trace}) we obtain:
\begin{equation}
 \sum_{\alpha=1}^m \mbox{tr} \left[ C^{(\alpha)\dagger} C^{(\alpha)} \right] -   \sum_{\beta=1}^n \mbox{tr} \left[ D^{(\beta)\dagger}  D^{(\beta)} \right] = N \label{main}
\end{equation} 
Define 
\begin{equation} 
\sum_{\alpha=1}^m C^{(\alpha)\dagger} C^{(\alpha)} = J \geq 0 \label{J}
\end{equation}
\begin{equation}
\sum_{\beta=1}^n D^{(\beta)\dagger}  D^{(\beta)} = K \geq 0. \label{K}
\end{equation}
Since $C^{(\alpha)\dagger}C^{(\alpha)}$ is Hermitian so is $J$. We can therefore perform a unitary transformation $U$ that diagonalizes $J$. By the trace condition the same unitary transformation automatically diagonalizes $K$. Then equation (\ref{trace}) becomes
\[ \tilde{J} - \tilde{K} = 1 \]
where $\tilde{J} = UJU^{\dagger}$ and  $\tilde{K} = UKU^{\dagger}$

Let the eigenvalues of $\tilde{J}$ be $j_i^2$ and those of $\tilde{K}$ be $k_i^2$  $(0 \leq i \leq N)$. Since $\tilde{J} = \mbox{diag}(j_1^2, j_2^2 \ldots j_N^2)$ and  $\tilde{K} = \mbox{diag}(k_1^2, k_2^2 \ldots k_N^2)$ we have the relation
\[ j_i^2 = k_i^2 + 1 \]
from the trace condition. Since $k_i^2 \geq 0$ the number of eigenvalues $j_i$ must be greater than or equal to the number of $k_i$, i.e. $m \geq n$ as mentioned earlier.

Define $\varphi_i$ so that
\[ j_i = \cosh \varphi_i \qquad, \qquad k_i = \sinh \varphi_i .\]
Now define

\begin{equation}
C^{(\alpha)} = [ \cosh \varphi ]\; M^{(\alpha)} 
\end{equation}
\begin{equation}
D^{(\beta)} = [ \sinh \varphi ] \; N^{(\beta)}
\end{equation}
where we have extracted the matrices $[\cosh \varphi]$ and $[\sinh \varphi]$ from the matrices $C^{(\alpha)} $ and $D^{(\beta)}$ respectively. It follows from equations (\ref{J}) and (\ref{K}) that
\begin{equation}
\sum_{\alpha=1}^m M^{(\alpha)\dagger} M^{(\alpha)} = 1\label{M}
\end{equation}
\begin{equation}
\sum_{\beta=1}^n N^{(\beta)\dagger}  N^{(\beta)} = 1 .\label{N}
\end{equation}
Parameterizing the matrices $M$ and $N$ are already known. In the cases where $\sinh \varphi = 0$, for all $\varphi$, we have only a smaller set of matrices to parameterize and this is identical to the case of having a completely positive map. Here we assume that $\sinh \varphi \neq 0$ and see how many parameters we need to write the map in the most general case (up to a unitary transformation). 

Since the matrices $M^{\dagger}M$ are hermitian we first choose a unitary transformation $W_1$ that can diagonalize $M^{(1)\dagger}M^{(1)}$:
\[ W_1^{\dagger} M^{(1)\dagger}M^{(1)} W_1 = 
\left( \begin{array}{cccc}
\cos^2 \theta_1^{(1)} & & & \mbox{\Large{0}} \\
 & \cos^2 \theta_2^{(1)} & &  \\
 & & \ddots & \\
\mbox{\Large{0}} & & &  \cos^2 \theta_N^{(1)}
\end{array}
\right)
\]

We can make further simplifications on $M^{(1)\dagger}M^{(1)}$ by noting the following \cite{ecg1}: We have at our disposal now $m$ matrices such that 
\begin{equation}
\sum_{\alpha=1}^{m}  W_1^{\dagger} M^{(\alpha)\dagger}M^{(\alpha)} W_1 =1, \label{Mnew}
\end{equation}
out of which the first term, $ W_1^{\dagger} M^{(1)\dagger}M^{(1)} W_1$ has been diagonalized. We can now make the following transformations on the first two terms of the sum in (\ref{Mnew}) without changing the sum itself or the eigenvalues of the map:
\begin{eqnarray*}
W_1^{\dagger} M^{(1)\dagger}M^{(1)} W_1 & \rightarrow  \\ 
& W_1^{\dagger} M^{(1)\dagger}M^{(1)} W_1 - \left( \begin{array}{cccc}
\cos^2 \theta_1^{(1)}-1 & & & \mbox{\Large{0}} \\
 & 0 & &  \\
 & & \ddots & \\
\mbox{\Large{0}} & & &  0
\end{array}
\right) \label{trans1}
\end{eqnarray*}
\begin{eqnarray*}
W_1^{\dagger} M^{(2)\dagger}M^{(2)} W_1 & \rightarrow  \\ 
& W_1^{\dagger} M^{(2)\dagger}M^{(2)} W_1 + \left( \begin{array}{cccc}
\cos^2 \theta_1^{(1)}-1 & & & \mbox{\Large{0}} \\
 & 0 & &  \\
 & & \ddots & \\
\mbox{\Large{0}} & & &  0
\end{array}
\right) \label{trans1}
\end{eqnarray*}
Such transformations which may be done on the set of matrices \{$M$\} (or on \{$N$\})do not change the sum in (\ref{M}) (or in (\ref{N})). The map which is made up of the {\em sums} in (\ref{M}) and (\ref{N}) hence remains unchanged and so these transformations allow us to fix $\cos \theta_1^{(1)}=1$. This freedom corresponds to performing orthogonal transformations (rotations) in the space containing the matrices $M^{(\alpha)}$. This in turn can be interpreted as an orthogonal transformation on the density matrix $\rho$ on which the map acts. An alternate way of looking at the transformation is to note that the sum of two matrices, $A + B$ can always be written as the sum of two other matrices say, for instance, $C+D$ where $C=(A+3B)/2$ and $D=(A-B)/2$. Once we have set $\cos \theta_1^{(1)} = 1 $, we parameterize $M^{(1)}$ using $N-1$ angles $\theta^{(1)}_i \; ; \; 2 \leq i \leq N$. 

Applying the transformation $W_1$ to both sides of equation (\ref{M}), exploiting the extra freedom mentioned above and after doing some re-labeling, we obtain
\begin{equation}
\sum_{\alpha=2}^m \tilde{M}_1^{(\alpha)\dagger} \tilde{M}_1^{(\alpha)} = 
\left( \begin{array}{cccc}
0 & & & \mbox{\Large{0}} \\
 & \sin^2 \theta_2^{(1)} & &  \\
 & & \ddots & \\
\mbox{\Large{0}} & & &  \sin^2 \theta_N^{(1)}
\end{array}
\right)
 \label{M1}
\end{equation}
 Where $\tilde{M}_1^{(\alpha)} = W_1^{\dagger} M^{(\alpha)} V_1$ for $2 \leq \alpha \leq m$. $V_1$ is another unitary matrix that reduces $\tilde{M}_1^{\alpha}$ to the form that we want. We can now focus on the set of $(N-1) \times  (N-1)$ matrices $M_1^{(\alpha)}$ defined as
\[ M_1^{(\alpha)} \equiv  
\left( \begin{array}{cccc}
 \sin \theta_2^{(1)} & & & \mbox{\Large{0}} \\
 & \sin \theta_3^{(1)} & &  \\
 & & \ddots & \\
\mbox{\Large{0}} & & &  \sin \theta_N^{(1)}
\end{array}
\right) \tilde{M}_1^{(\alpha)} \quad ; \quad  2 \leq \alpha \leq N
 \label{M1a}
\]
where we have dropped the first row and column of $ \tilde{M}_1^{(\alpha)}$ on the right hand side of the equation and also extracted the factor containing $\sin \theta_i^{(i)}$ from it. We assume that none of the $\sin \theta_i^{(i)}$ are zero since we are interested in computing the maximum number of parameters required for describing a generic extremal map that is not completely positive.

From equation (\ref{M1}) it follows that
\[ \sum_{\alpha =2}^{m} M_1^{(\alpha)\dagger} M_1^{(\alpha)} = 1_{(N-1) \times (N-1)} \]
the matrix $ M_1^{(2)}$ in the first term of this sum can be parameterized using $N-2$ parameters using exactly the same procedure as before. Using a unitary transformation $W_2$ and a further orthogonal transformation (if needed) we can transform $ M_1^{(2)}$ to the following form as before
\[  M_1^{(2)} \longrightarrow
\left( \begin{array}{cccc}
 1 & & & \mbox{\Large{0}} \\
 & \cos \theta_2^{(2)} & &  \\
 & & \ddots & \\
\mbox{\Large{0}} & & &  \cos \theta_{N-1}^{(2)}
\end{array}
\right)
 \label{M2}
\]
Repeating this procedure $m$ times, we parameterize all the matrices $C^{(\alpha)}$. $D^{(\beta)}$ can also be parameterized in the same fashion. The total number of parameters needed can be computed as follows. There are $N$ angles $\varphi_i$. To parameterize the matrices $M^{(\alpha)}$ we need $(N-1) + (N-2) + \ldots (N-m)$ parameters and for $N^{(\beta)}$ we need $(N-1) + (N-2) + \ldots (N-n)$ parameters. So in total we need
\begin{equation}
 N^2 - \frac{m(m-1)+n(n-1)}{2} \label{number}
\end{equation}
parameters.

Note that the matrices $C^{(\alpha)}$ and  $D^{(\beta)}$ are determined only up to $m+n$ unitary matrices according to
\[ C^{(\alpha)} \longrightarrow  C^{(\alpha)}U^{(\alpha)} \]
which leave equation (\ref{main}) unchanged. We can see this also in the manner we defined $\tilde{M}_1^{(\alpha)} = W_1^{\dagger} M^{(\alpha)} V_1$ where we had to introduce the arbitrary unitary matrix $V_1$. 

\section{Dynamical Maps as Contractions}\label{sec3}

A straightforward way of generating positive maps is to consider the unitary evolution of two systems coupled to each other. Let $S$ be the system of of interest and $R$ the second system. Let the dimensionality of $S$ be $d$ and that of $R$ be $N$. $R$ can be treated as a `reservoir' with which $S$ is interacting. Dynamical maps representing the time evolution of $S$ can then  be thought of as contractions on the unitary evolution of the combined system. If we choose a direct product density matrix as the initial state then the dynamics of the coupled system is given by
\[ {\cal{R}} = \rho_S \times \tau_R \longrightarrow V \rho_S \times \tau_R V^{\dagger} \]
where $V$ is a unitary matrix in the direct product space ${\cal{H}_S \times \cal{H}_R}$ 
Using the index notation employed in the previous discussion
\begin{equation}
\rho_{rs} \times \tau_{ab} \longrightarrow V_{ra; r'a'} \rho_{r's'} \tau_{a'b'} V^*_{sb; s'b'}. 
\end{equation}
The evolution of the system $S$ is extracted using the partial trace operation which  is a contraction.
\[ \rho_{rs} \longrightarrow {\mbox{tr}}_R \left( V_{ra; r'a'} \rho_{r's'} \tau_{a'b'} V^*_{sb; s'b'} \right) =  V_{rn; r'a'} \rho_{r's'} \tau_{a'b'} V^*_{sn; s'b'}. \]
For simplicity we assume that $\tau$ can be made diagonal by a suitable unitary transformation in ${\cal{H}}_R$ with eigenvalues $\tau(1), \tau(2) \ldots \tau(n)$. Then the map on $S$ is
\begin{equation}
\rho_{rs} \longrightarrow \sum_{\nu,n, r',s'} V_{rr'}(n, \nu) \rho_{r's'}\tau(\nu) V^*_{s's}(n,\nu) . \label{contraction}
\end{equation}
Here the operator $V$ has been rewritten in a manner suggestive of the form of a completely positive map that is not extremal, given in (\ref{notextremal}) . i.e.
\[ \rho \longrightarrow \sum_{\nu} \sum_{\alpha} k(\nu)C^{(\alpha)}_{\nu} \rho C^{(\alpha)\dagger}_{\nu}. \]
To get the form of the map in (\ref{notextremal}) it is sufficient that the dimensionality $N$ of the reservoir to be the same as that of the system. i.e. $N=d$. With this restriction $\tau(\nu)$ could correspond to a mixed state.  If we further restrict $\tau$ to correspond to a pure state so that it has only one eigenvalue then the  map is extremal and reduces to the standard form (\ref{completely})
\[ \rho_{rs} = \sum_n V_{rr}^{(n)} \rho_{r's'} V_{ss'}^{(n)*} \simeq \sum_{\alpha}C^{(\alpha)}\rho C^{(\alpha) \dagger} \]
in which $\alpha$ runs over $1 \leq \alpha \leq d$. In other words extremal completely positive maps are contractions of unitary evolution in a space in which the system is coupled to a reservoir, of the same number of dimensions of the system, whose initial state is a pure projection. 

We note here that all these maps are completely positive maps (not necessarily extremal). if the dimension of the reservoir is made bigger than or equal to $d^2$ then any map on $S$ (not necessarily extremal) can be expressed as a contraction of the unitary evolution in a space in which the system is coupled to a reservoir whose initial state is a {\em pure projection}. 

We can also carry out the inverse construction where we start with a completely positive map and view it as a unitary transformation on a larger system. Given an extremal completely positive map of the form (\ref{completely}), we can construct a unitary matrix $V$ in $mn$ dimensions with:
\[ V_{r\alpha; r'1} = C_{rr'}^{(\alpha)}. \]
The conditions on $C^{(\alpha)}$ are transcribed into
\[ \sum_{\alpha, n} V^*_{r\alpha; n1} V_{s\alpha;n1} = \delta_{rs}\]
which is necessary for $V$ to be a unitary matrix. The ambiguity in constructing the other elements of $V$, where the last index is not equal to 1, does not affect the map. In short, $V$ can be constructed in such a fashion that it corresponds to any given dynamical map on the system along with a particular choice of the states $\tau$ of the reservoir. 

\section{Not Completely Positive Maps as Contractions}\label{sec4}

What about not completely positive maps? To obtain such a map as a contraction we have to generalize the auxiliary space ${\cal{H}}_R$ to be a space with an indefinite metric and $V$ to be a pseudo unitary operator in the $mn$ dimensional space. Positivity of the map is guaranteed if the generalized density matrix of the extended system is initially entirely within the convex set of positive metric states of the $mn$ dimensional space. The sum over the index $n$ in (\ref{contraction}) goes over both positive and negative metric terms; but the resultant density matrix is nonnegative. In general the not completely positive maps have to be viewed as contractions of the evolution of unphysical systems. For example the complex conjugation of the density matrix $\rho$ of the system is a positive map which is not completely positive. Physically it has the meaning of time reversal of the system. However if we view it as a contraction of the evolution of two coupled systems, then this corresponds to time reversal of only of the systems which is rather meaningless. 

We can invert this derivation to realize the most general not completely positive map as the contraction of a larger evolution in an indefinite metric space for the reservoir. To make the map extremal we further restrict the density matrix of the reservoir $\tau$ to have a single eigenvector(with positive metric) with eigenvalue unity. Since such reservoirs are somewhat artificial, we have to consider this reconstruction as a purely formal device.

\section{Summary}

We have studied linear dynamical maps which take the set of density matrices into the set of density matrices. These maps form a convex set which is also compact in the case of completely positive maps. It has already been shown in \cite{ecg3} that a completely positive extremal map contain at most $N$ terms, requiring a total of $N(N-1)/2$ terms to parameterize each of the terms up to a set of $N \times N$ unitary matrices. The completely positive maps can be viewed as the contraction of unitary evolution in an extended space. Extremal maps correspond to the case where the auxiliary system in the extended space is a pure projection. Conversely we can reconstruct the unitary evolution of the expanded system from the map itself. 

These considerations are extended in this paper to positive but not completely positive maps. The extremal maps again have at most $N$ terms. The number of parameters required to describe each one of these terms up to a set of unitary $N \times N$ transformations is given in equation (\ref{number}). We can obtain these maps also as contraction of an extended system. But here the extended system has a pseudo unitary evolution matrix. It is also possible to obtain this pseudo unitary evolution starting from the maps.

These results generalize the results obtained two decades ago by Gorini and Sudarshan\cite{gorini1} for $2 \times 2$ matrices. 

Needless to say, however complicated the dynamical processes leading to the linear stochastic evolution that is represented by the dynamical map, we see that the same dynamics obtains when we couple the system to a reservoir having dimension $N^2$. In the case of an extremal map it suffices to couple the system to a reservoir of dimension $N$. 

In this paper we have dealt only with dynamical maps and not with the continuous semigroup of evolution. This study was carried out by Kossakowski\cite{kossakowski} and followed by others\cite{lindblad, gorini3}. In these, while the semigroup generators are parameterized no attempt is made to embed them in a larger system. Since the Zeno effect\cite{misra} operates for very small time intervals, care must be taken in generating a semigroup from the dynamics of an extended system. We hope to examine this question in the near future.


\begin{thebibliography}{99}
\bibitem{com2} This was already recognized in reference \cite{ecg1}. See also reference \cite{lindblad}.
\end{thebibliography}
\end{document}